\documentstyle[12pt]{article}
\begin{document}
\author{A.A.Zhmudsky}
\date{\today}
\title{Electromagnetic radiation of the travelling spin wave propagating in
an antiferromagnetic plate. Exact solution.}
\maketitle
\begin{abstract}
The exact solution of radiation problem of a spin wave travelling in an
antiferromagnetic (AFM) plate was found. The spin wave
with in-plane oscillations of antiferromagnetism vector was considered. In
this case the magnetization vector is oscillating being perpendicular to
the AFM plate and depends on time and plane coordinates as travelling wave
does. This model allows to obtain exact analytical expression for Hertzian
vector and, consequently, the retarded potentials and field strengths as
well.

It is shown that expressions obtained describe Cherenkov radiation caused
by the travelling wave. The radiated electromagnetic wave is the $TEM$ type
if a phase velocity exceeds the speed of light. Otherwise electric
and magnetic field values exponentially decrease in the direction normal to
the plate. The energy losses were evaluated also.
\end{abstract}
\section{Introduction} It is known that not only particles can be the
Cherenkov radiation sources but also, so-called, superlight "reflections"
\cite{BG} formed by the motion of particles number large enough.

We want to point out that such effects can be observed at spin wave
propagation on the magnetic surface. In particular, it is possible to
receive the exact solution of radiation problem of a spin wave travelling
in an antiferromagnetic plate.

\section{Spin wave propagation in the antiferromagnetic plate}
Let us consider planar antiferromagnetic (AFM) containing two magnetic
sublattices with magnetizations $\vec M_1$ and $\vec M_2$. Total
magnetization of the AFM $\vec M=\vec M_1+\vec M_2$ in the ground
state is equal zero ($\vec M_1=-\vec M_2$, $|\vec M_1|=|\vec M_2|=M_0$).

We will use the $\sigma-$model approach based on the equation for the
antiferromagnetism vector $\vec l=(\vec M_1-\vec M_2)/2M_0$ \cite{BI,AM,Mik}.
The effective Lagrangian of the $\sigma-$model for the $\vec l$ vector
reads as \cite{IK}:
\begin{equation}
{\cal L}=\frac {\alpha M_0^2}2\int\limits_{}^{}\left\{\frac 1{c^2}\left(
{\partial\vec l\over\partial t}\right)^2-(\nabla\vec l)^2-w(\vec l)
\right\}d^2x,
\label{eq:dip01}\end{equation}
where $w(\vec l)=\frac 12\beta_1 l^2_y+\frac 12\beta_2 l^2_z$
($0<\beta_1<\beta_2$) is anisotropy energy,
$c=\gamma M_0\sqrt{ \alpha\delta/2}$
\footnote{In the simplest case of zero field and zero Dzyaloshinakii
interaction.}. The phenomenological constants $\delta$ and $\alpha$
describe the homogeneous and inhomogeneous exchange interactions,
respectively.

The dynamic equations for $\vec l$ can be written as Euler-Lagrange equations
for the Lagrangian (\ref{eq:dip01}). Using ${\vec l}^2=1$
these equations may be presented in the form \cite{IK}:
\begin{equation}
\left[\vec l\times{\delta{\cal L}\over\delta\vec l}\right]=0.
\label{eq:dip02}\end{equation}
For the planar AFM it is convenient to represent the dynamics of unit vector
$\vec l$ by means of the angular variables:
\begin{equation}
l_3=\cos\theta, \qquad l_1+il_2=\sin\theta\exp(i\varphi);
\label{eq:dip03}\end{equation}
where the polar axis is directed along the easy axis of the AFM. The equations
of motion for $\theta$ and $\varphi$ can be written in the form:
\begin{eqnarray}
\alpha\left(\nabla^2\theta-\frac 1{c^2}{\partial^2\theta\over\partial t^2}
\right)+
\alpha\sin\theta\cos\theta\left[(\nabla\varphi)^2-\frac 1{c^2}
\left({\partial\varphi\over\partial t}\right)\right]-
{\partial w_a\over\partial\theta}=0, \nonumber \\
\alpha\nabla(\sin^2\theta\nabla\varphi)-\frac {\alpha}{c^2}
{\partial\over\partial t}
\left(\sin^2\theta{\partial\varphi\over\partial t}\right)-
{\partial w_a\over\partial\varphi}=0.
\label{eq:dip04}\end{eqnarray}
We will search the solution like travelling wave which propagate in the
$X0Y$ plane ($\theta=\pi/2$) and $\varphi=\varphi((\vec k\vec r-\omega
t)/k\sqrt{\alpha})$. In this the case equations (\ref{eq:dip04}) give:
\begin{equation}
\left(\frac {V^2}{c^2}-1\right){\partial^2\varphi\over\partial\xi^2}+
\sin\varphi\cos\varphi=0,
\label{eq:dip05}\end{equation}
where $V=\omega/k$, $\vec k$ - wave vector and $\vec r$ radius-vector in
the magnetic plane, $\omega$ - frequency.

At small $k$ phase velosity of spin wave increases infinitely that is why
$V>c$, equation (\ref{eq:dip05}) has stable harmonic solution and Cherenkov
radiation takes place.

Vectors $\vec l$ and $\dot{\vec l}$ oscillate in the AFM plane, thus the
magnetization vector $\vec M\sim [\vec l\times\dot{\vec l}]$ is normal to
the AFM plane. Evidently, if $\varphi\ll 1$ the dependence of magnetization
vector from time and space variable becomes:
\begin{equation}
\vec M=\vec M_0\exp(-i\omega t+i\vec k\vec r),
\label{eq:dip06}\end{equation}
where $\vec M_0=(0,0,M_0)$, $\omega$ - frequency and $\vec k$ wave vector
of the travelling wave.

In the following (next) section we will show that dependence
(\ref{eq:dip06}) allows to receive the exact solution of the radiation
problem.

\section{Exact solution of the Cherenkov radiation problem of a spin wave}
At a given functional dependence of magnet dipole moment $\vec M$ on the
space and time variable the exact solution of D'Alemberian equation for the
magnetic Hertzian vector $\vec\Pi_m$:

\begin{equation}  \Delta\vec\Pi_m-{1\over c^2}
{\partial^2\vec\Pi_m\over\partial t^2}=-4\pi\vec\Pi_m,
\label{eq:dip1}\end{equation}
is expressed by the retarded potential:
\begin{equation}  \vec \Pi_m={1\over c}\int\limits_V
{\vec M(\vec r',t-{\displaystyle{|\vec r-\vec r'|\over c}})dV'\over
|\vec r-\vec r'|},
\label{eq:dip2}\end{equation}
where $\vec M$ is a magnetic dipole moment per area unit, $\vec r$ is the
distance between the origin and the observation point, $\vec r'$ is the
distance between the origin and the source point where the magnet dipole
moment element is placed.

Carrying out the integration (\ref{eq:dip2}) one can find the vector
potential $\vec A$ and scalar potential $\varphi$ via:
\begin{equation} \vec A=rot\vec\Pi_m,\qquad \varphi=0.
\label{eq:dip3}\end{equation}
And the expression  of electric and magnetic fields as usual:
\begin{eqnarray}
\vec E&=&-{1\over c}{\partial\over\partial t}rot\vec\Pi_m\nonumber\\
\vec H&=&rot\,rot\;\vec\Pi_m .
\label{eq:dip4}\end{eqnarray}
It is easy to test that Lorentz gauge is satisfied identically at the made
definitions.

Taking into account that a magnet moment in (\ref{eq:dip2}) has to contain a
retarded time, we shall obtain:
\begin{eqnarray}
\vec \Pi_m&=&{\vec M_0\over c}\int\limits_V
{\exp(-i\omega t+{\displaystyle{i\omega\over c}}|\vec r-\vec r'|+ik_xx')dV'
\over |\vec r-\vec r'|} \nonumber \\
&=&{\vec M_0\cdot\exp(-i\omega t+ik_xx)\over c}\cdot\nonumber \\
& &\int\limits_{-\infty}^{\infty}
\int\limits_{-\infty}^{\infty}
{\exp({\displaystyle{i\omega\over c}}\sqrt{(x-x')^2+(y-y')^2+z^2}+
ik_x (x'-x))dx'dy'
\over \sqrt{(x-x')^2+(y-y')^2+z^2}}\nonumber \\
\label{eq:dip5}\end{eqnarray}
We assume $z-$axis to be directed along the normal to the plate and
$x-$axis coincides with the wave vector $\vec k$. Evidently, that $\vec
k\vec r=kx$.

Let's use the polar coordinates $x'-x=r\cos \varphi$ and $y'-y=r\sin
\varphi$ for the further calculation. This gives:
\begin{eqnarray} \vec \Pi_m&=&{\vec M_0\exp(-i\omega t+ikx)\over c}
\cdot\int\limits_0^{\infty}\int\limits_0^{2\pi}
{\exp({\displaystyle{i\omega\over c}}\sqrt{r^2+z^2})\exp(ikr\cos\varphi)
rdrd\varphi\over
\sqrt{r^2+z^2}}\nonumber\\
&=&{\vec M_0\exp(-i\omega t+ikx)\over c}\int\limits_0^{\infty}
\int\limits_0^{2\pi}
\exp(ikr\cos\varphi)d\varphi {\exp({\displaystyle{i\omega\over c}}
\sqrt{r^2+z^2})rdr\over\sqrt{r^2+z^2}}\nonumber\\
&=&2\pi{\vec M_0\exp(-i\omega t+ikx)\over c}\int\limits_0^{\infty}
J_0(kr){\exp({\displaystyle{i\omega\over c}}\sqrt{r^2+z^2})rdr\over
\sqrt{r^2+z^2}},
\label{eq:dip6}\end{eqnarray}
where $J_0(kr)$ is the Bessel function of the first kind of zero order. In
(\ref{eq:dip6}) the well-known equality for Bessel functions was used (e.g.
9.1.21 in \cite{Abramowitz}).

Let's use the Euler's formula and represent the exponential form
(\ref{eq:dip6}) in the trigonometric one (by sine and cosine):
\begin{eqnarray}
\vec \Pi_m&=&2\pi{\vec M_0\exp(-i\omega t+ikx)\over c}\int
\limits_0^{\infty}
J_0(kr){\cos({\displaystyle{\omega\over c}}\sqrt{r^2+z^2})rdr
\over\sqrt{r^2+z^2}}\nonumber\\
&+&2\pi i{\vec M_0\exp(-i\omega t+ikx)\over c}\int\limits_0^{\infty}
J_0(kr){\sin({\displaystyle{\omega\over c}}\sqrt{r^2+z^2})rdr
\over\sqrt{r^2+z^2}}
\label{eq:dip7}\end{eqnarray}
Each integral from (\ref{eq:dip7}) can be evaluated exactly (not
approximately), by virtue \cite[p. 775, (6.737)]{Ruzhik}:
\begin{equation}
\int\limits_0^{\infty}J_0(kr){\cos({\displaystyle{\omega\over c}}
\sqrt{r^2+z^2})rdr
\over \sqrt{r^2+z^2}}=
\cases{{\displaystyle-\sqrt{\pi z\over 2}{\displaystyle{N_{-\frac12}(|z|
\sqrt{{\omega^2\over c^2}-k^2})
\over \sqrt[4]{{\omega^2\over c^2}-k^2}}},|{\omega\over c}|>k;}\cr
{\displaystyle\sqrt{2z\over\pi}{\displaystyle{K_{\frac12}(|z|
\sqrt{k^2-{\omega^2\over c^2}})
\over \sqrt[4]{k^2-{\omega^2\over c^2}}}},|{\omega\over c}|<k.} }
\label{eq:dip8}\end{equation}
\begin{equation}
\int\limits_0^{\infty}J_0(kr){\sin({\displaystyle{\omega\over c}}
\sqrt{r^2+z^2})rdr
\over\sqrt{r^2+z^2}}=\cases
{{\displaystyle\sqrt{\pi z\over 2}{\displaystyle{J_{-\frac12}
(|z|\sqrt{{\omega^2\over c^2}-k^2})
\over \sqrt[4]{{\omega^2\over c^2}-k^2}}},|{\omega\over c}|>k;}\cr
{\displaystyle 0,|{\omega\over c}|<k.}}
\label{eq:dip9}\end{equation}
Bessel function of half-integer order may be expressed through elementary
functions (\cite{Abramowitz}):
\begin{eqnarray}
N_{-\frac12}(x)&=&\sqrt{2\over \pi x}\sin(x),\qquad
K_{\pm\frac12}(x)=\sqrt{\pi\over 2x}\exp(-x)\nonumber \\
J_{-\frac12}(x)&=&\sqrt{2\over \pi x}\cos(x),
\label{eq:dip10}\end{eqnarray}
where $N_{-\frac12}(x)$ and $K_{\pm\frac12}(x)$ are modified Bessel's
functions of half-integer order. It is convenient to make the following
definition:
\begin{equation}
q=\cases{\sqrt{{\omega^2\over c^2}-k^2},\qquad |{\omega\over c}|>k \cr
         i\sqrt{k^2-{\omega^2\over c^2}},\qquad |{\omega\over c}|<k}
\label{Q}\end{equation}
One can return to the exponents and with respect to (\ref{eq:dip10}) express
(\ref{eq:dip7}) in the form:
\begin{equation}
\vec \Pi_m=\frac iq\frac {2\pi\vec M_0}c\exp(-i\omega t+ikx+iq|z|)
\label{eq:dip11}\end{equation}
Expression (\ref{eq:dip11}) describes typical case of Cherenkov radiation.
If wave velocity exceeds the light one then the $TEM$ wave is radiated. In
the opposite case electromagnetic fields exponentially decrease in
$z-$direction.

Using (\ref{eq:dip4}) we readily find the electric and magnetic fields
expression:
\begin{eqnarray}
\vec E=\vec e_2\frac iq\frac {2\pi\vec M_0\omega k}c^2\exp(-i\omega t+
ikx+iq|z|) \nonumber\\
\vec H=i\frac kq\frac {2\pi\vec M_0}c\exp(-i\omega t+ikx+iq|z|)\left\{
\mp q\vec e_1+k \vec e_3\right\},
\label{eq:dip12}\end{eqnarray}
where $\vec e_1, \vec e_2, \vec e_3$ - unit Cartesian vectors, signs $\mp$
correspond to the $z>0$ and $z<0$ respectively.

Also easy the Pointing vector can be written:
\begin{equation}
\vec S={\pi\omega M_0^2 k^2\over q^2c^2} \left\{\pm q\vec e_3+k \vec
e_1\right\},
\label{eq:dip13}\end{equation}
where signs $\pm$ correspond to the $z>0$ and $z<0$ respectively.

\section{Conclusion}
Evidently, that the travelling wave of the electric dipoles at the plate like
(\ref{eq:dip06}):
$$ \vec P=\vec P_0\exp(-i\omega t+i\vec k\vec r),$$
gives the same solution as (\ref{eq:dip11}) with the simple changes
$\vec\Pi_m\to \vec\Pi_e$ and $\vec M_0\to \vec P_0$ and corresponding
expressions for fields: $$\vec A={1\over c}{\partial\vec\Pi_e\over\partial
t},\qquad \varphi = -div\vec\Pi_e$$

\section{Acknowledgments}
I am grateful to Dr. Boris Ivanov for helpful discussion and advice.

\end{document}